\documentclass[twocolumn,twoside,slac]{revtex4}
\usepackage{graphicx}
\usepackage{fancyhdr}
\begin{document}

\title{Sharing a conceptual model of grid resources and services}

%

\author{Sergio Andreozzi}\thanks{Work supported by the EU DataTAG project}
\affiliation{INFN-CNAF, Bologna, I-40126, ITALY}
\author{Massimo Sgaravatto}\thanks{Work supported by the EU DataGrid project}
\affiliation{INFN-PD, Padova, I-35131, ITALY}
\author{Cristina Vistoli}\thanks{Work supported by the EU DataTAG project}
\affiliation{INFN-CNAF, Bologna, I-40126, ITALY}

\begin{abstract}

Grid technologies aim at enabling a coordinated resource-sharing
and problem-solving capabilities over local and wide area networks
and span locations, organizations, machine architectures and
software boundaries. The heterogeneity of involved resources and
the need for interoperability among different grid middlewares
require the sharing of a common information model. Abstractions of
different flavors of resources and services and conceptual schemas
of domain specific entities require a collaboration effort in
order to enable a coherent information services cooperation.

With this paper, we present the result of our experience in grid
resources and services modelling carried out within the Grid
Laboratory Uniform Environment (GLUE) effort, a joint US and EU
High Energy Physics projects collaboration towards grid
interoperability. The first implementation-neutral agreement on
services such as batch computing and storage manager, resources
such as the hierarchy cluster, sub-cluster, host and the storage
library are presented. Design guidelines and operational results
are depicted together with open issues and future evolutions.

\end{abstract}

\maketitle

\thispagestyle{fancy}


\section{INTRODUCTION}

Grid technologies aim at enabling a coordinated resource-sharing
and problem-solving capabilities over local and wide area networks
and span locations, organizations, machine architectures and
software boundaries. The heterogeneity of involved resources and
the need for interoperability among different grid middleware
solutions require the sharing of a common information model in
order to enable both intra- and inter-grid resources awareness.

The research area of computing in High Energy and Nuclear Physics
(HENP) is populated by several Grid related projects that mostly
rely on basic services provided by the Globus
Toolkit~\cite{GLOBUSTK} and the Condor Project~\cite{CONDOR}. Due
to the large adoption of the provided functionalities,
interoperability issues are mostly related to what is built on top
of these components. For the purpose of enabling HENP Grid
middlewares interoperability, the Grid Laboratory Uniform
Environment (GLUE) collaboration~\cite{GLUEEFFORT}, a joint US and
EU High Energy Physics projects effort, has been set up.

One of the main achievements of this collaboration has been
carried out in the context of the GLUE Schema activity. The main
purpose was to define a common resource information model to be
used as a base for Grid Information Service (GIS) for both
resource discovery and monitoring activities. Starting from the
Globus MDS schema~\cite{MDSSchema} and the EU DataGrid (EDG)
schema, the first implementation-neutral agreement on services
(such as batch computing and storage managers) and systems (such
as the hierarchy cluster, sub-cluster, host and the storage
library) has been defined~\cite{GLUESchema}. The EU DataTAG (EDT)
project~\cite{EDT} has contributed in the collection of
requirements from several projects (mainly EDG) and has developed
both the implementation-neutral description by means of Unified
Modeling Language (UML)~\cite{UML} class diagram and schema
implementation for the LDAP data model~\cite{GLUE4LDAP}.

In this paper, we recall the outcomes of the collaboration to
which we have participated, and we also suggest refinements and
improvements based on our experience related to both analysis,
implementation and deployment within EDG~\cite{EDG},
EDT~\cite{EDT} and LCG~\cite{LCG} testbeds.

We present the involved entities categorized into two main
categories:
\begin{itemize}
    \item System: a set of connected items or devices which operate together as a functional
whole.
    \item Service: actions that form a coherent whole from the point of view of service providers and service
requesters.
\end{itemize}

From the viewpoint of discovery and monitoring, the distinction
between systems and services is fundamental, since even though
they are strongly related (systems provide services), they have
different life-cycle and different status related attributes. The
main focus was on the service level in order to enable an
efficient service selection. Recently, within the DataTAG project,
extensions for the host system have been done in order to improve
monitoring capabilities.

This paper is organized as follow: in section~\ref{sec:SYSTEM} we
describe entities within the system category, while in
section~\ref{sec:SERVICE} services are discussed. In both cases,
defined concepts are recalled and feedback from our deployment
experience is described. In section~\ref{sec:IMPL}, the
implementation results are presented, while in
section~\ref{sec:RELATED} related works are mentioned and compared
to the GLUE Schema. Finally, in section~\ref{sec:CONCLUSION}
conclusions and plans for future work are depicted.


\section{Modelling systems\label{sec:SYSTEM}}

As mentioned in the introduction, a system is here defined as a
set of connected items or devices which operate together as a
functional whole. Within the GLUE Schema, two  main systems
categories have been defined: cluster systems providing computing
services, and storage systems providing storage spaces.

\subsection{Cluster Systems}

A cluster is essentially a container that groups together
computing nodes (hosts), such as a computing farm. Since in the
context of computing in HENP, a cluster is composed by many nodes,
in order to avoid poor performance in the resource discovery
process, the concept of subcluster was introduced. A subcluster
represents `homogeneous' collection of computing nodes, where the
homogeneity is defined by a collection whose some node attributes
(which can be freely selected among the ones defined for the
single node entity) all have the same value. For example, a
subcluster could represent a set of nodes with the same
architecture, operating system, CPU model, etc. It must be
stressed that the elements of a subcluster are homegeneous only
with respect to the considered chosen attributes. Subclusters
therefore provide a convenient way of representing collection of
nodes, useful in the resource discovery process. The host
(computing node) element represents detailed information (related
to both hardware and software) of a specific node.

In summary, a cluster is a set of nodes, and nodes can be
partitioned in disjointed sets called subclusters, for which a
summary description is available.

From our point of view, the cluster definition should be refined
enforcing the property of being a system. This implies that some
functionality is provided. In the cluster case, the provided
functionality is the ability of executing jobs. Therefore, all
nodes managed by the same batch system, form a unique cluster.

\subsection{Storage Systems}

In a Grid environment, storage systems can vary in complexity from
a single disk server to hierarchical massive storage systems.
Within the first phase of the GLUE Schema activity, the main goal
was mainly oriented towards the service component modelling (see
section ~\ref{SMS}), more than the system component. The modelled
storage system is called Storage Library and represents the
machine providing for the storage manager service. This entity
presents the file system component offered to the service, an
architecture component and a performance component.

Our opinion is that this concept should be refined. As is for the
cluster system, a storage system should allow the representation
of all entity participating in the service.

\section{Modelling services\label{sec:SERVICE}}

A service can be defined as an activity that performs some task.
Within core grid services, the computing service and the storage
manager service have been defined. Each modelled service has a
unique identifier, a human-readable name, a set of policies, a set
of access rights and a state.

Referring to the access rights, one of the main design guideline
was to move from the current practice of user-grained access right
to the virtual organization-grained one. This approach is
beneficial for the activity of authorization management in a
distributed environment such as the Grid. The idea is that virtual
organizations set up agreements with service providers. When a
service is requested by a user, both his membership and his
capabilities are verified using organization-based authorization
services. Considering this approach, local resources can therefore
avoid to maintain and publish the list of authorized user
identities. The two main proposals in this area are the Virtual
Organization Membership Service (VOMS)~\cite{VOMS} developed in
the context of the EDG-EDT collaboration, and the Community
Authorization Service (CAS)~\cite{CAS} developed by the Globus
Project.

\subsection{Computing Service}

As computing service, we identify a service able to provide
computing power to an application with a certain quality. Within
the GLUE Schema, the modelled service is called Computing Element
(CE) and it is a one-to-one mapping to an entry point into a batch
queueing system. Essentially a Computing Element represents a
queue of a local resource management system, such as PBS or LSF.
Since it is a service, it presents policy, state and access rights
attributes.

The CE concept was already present in the EDG Schema. With the
GLUE Schema the separation between system and service related info
was introduced.

In order to be able to perform a proper service selection, during
the matchmaking process some data related to the system providing
the service is needed (e.g. hosting operating system, available
software packages). Moreover, the desired data should be provided
as an aggregate description of the system part that can
participate in the service functionality (e.g. in a cluster, only
a subset of nodes can be assigned to a computing service). Such
flexibility is not well modelled at present.

\subsection{Storage Manager Service\label{SMS}}

With Storage service we identify a service which task is the
management of storage extents. With Storage Space, we identify a
storage extent managed under a uniform set of policies and having
the same access rights. Stored files can be accessed by means of
Data Access Protocols (e.g. GridFTP, rfio).

Currently, the Storage Service has been modelled as a
generalization of:
\begin{itemize}
    \item trivial file system
    \item Storage Resource Manager (SRM)~\cite{SRM}
    \item EDG Storage Element (SE)~\cite{SE}
\end{itemize}

Current practice shows that there is lack of some information. For
instance, a Storage Space is lacking of ownership info and of a
unique ID, possibly in the form of a URI.

\section{Implementations\label{sec:IMPL}}

Currently, the implementation-neutral description of the GLUE
Schema has been mapped into three different data models:
\begin{itemize}
    \item LDAP data model~\cite{GLUE4LDAP}
    \item Relational data model~\cite{GLUE4RGMA}
    \item XML data model~\cite{GLUE4XML}
\end{itemize}

The LDAP implementation has been done within the DataTAG project
and covers the full schema. The testing phase has been carried out
within the DataTAG testbed based on the EU DataGrid middleware
(release 1.4.x with Globus MDS 2.x). Strong support has been given
for the rewriting of both information providers and EDG Broker
interaction with the MDS. The schema implementation is going to be
deployed as the base schema in the next release of EDG, LCG and
VDT grid middlewares. This has been also contributed by
INFN~\cite{INFN} to the Globus Project under the signed Globus
Contributor's License.

The Relational implementation has been developed by DataGrid WP3
in the R-GMA~\cite{RGMA}.

The XML implementation has been developed by the Globus Project in
Globus Toolkit 3.0 (at present, only cluster system and computing
element).

\section{Related work\label{sec:RELATED}}

The first resource information model introduced within the HENP
Grid community was the one coming with the Globus MDS. This was
mainly oriented at modelling computer systems and lacked in
capabilities of modelling grid services. Conversely, the former
DataGrid schema was mainly oriented at creating abstractions for
grid services. While this presented new concepts useful for
discovery and matchmaking purposes, it did not clearly separates
systems from services.

Another work to mention is the NorduGrid information
model~\cite{NorduGrid}. The distinguish feature is the User
entity, modelled in order to provide per-user information, such as
available storage space and processors. This is a different
approach than the one taken within the GLUE Schema, where general
authorization rules are on a per-organization base.

It is important to mention also the outcomes of the ongoing
activities of the CIM Grid Schema Working Group of the Global Grid
Forum (GGF CGS-WG)~\cite{GGF_CGSWG}. The considered approach is
not only targeted at discovery and monitoring purposes, but also
at the more complex task of resource management. The chosen
strategy is to extend the industry standard Common Information
Model (CIM)~\cite{CIM}. While this approach provides detailed
resource description and relationship, it needs special framework
in order to offer the management interface. For the purpose of
distributed discovery, it needs to be interfaced with the Grid
Information Service. Due to the wider spectrum of goals envisioned
by CIM, the information model is more complex.

\section{Conclusions and future work\label{sec:CONCLUSION}}

In this paper we presented the results and the outcomes of the
GLUE Schema activity. Besides presenting the modelled entities,
both at the system and at the service level, the current
shortcomings and some ideas to address them have also been
described.

In the next future the focus will be on the the refinement of both
subcluster and storage library entities. We also envision the
evolution of the current proposal in order to model a general
ancestor service. Moreover, attention will be given to monitoring
requirements.

\begin{acknowledgments}
The authors wish to thank all the persons that have participated
to the GLUE Schema activity.
\end{acknowledgments}


\end{document}